\documentclass[letterpaper, 10 pt, conference]{ieeeconf}  

\IEEEoverridecommandlockouts                              

\usepackage{graphics} 
\let\labelindent\relax
\usepackage{xcolor}
\usepackage{framed,enumitem} 
\usepackage{epsfig} 
\usepackage{mathptmx} 
\usepackage{times} 
\usepackage{amsmath} 
\usepackage{amssymb}  
\usepackage{arydshln}
\usepackage{amsfonts}
\usepackage{newtxtext, newtxmath}

\usepackage[noadjust]{cite}
\title{\LARGE \bf
On modal observers for beyond rigid body $H_\infty$ control\\ in high-precision mechatronics*
}

\author{Yorick Broens, Hans Butler and Roland T\' oth
\thanks{*This work has received funding from the ECSEL Joint Undertaking (JU) under grant agreement No 875999 and from the European Union within the framework of the National Laboratory for Autonomous Systems (RRF-2.3.1-21.2022-00002).}
\thanks{Y.Broens, H.Butler and R.T\'oth are with the Department of Electrical Engineering, Eindhoven University of Technology, Eindhoven, The Netherlands. H.Butler is also affiliated with ASML, Veldhoven, The Netherlands. R.T\'oth is also affiliated with the Systems and Control Laboratory, Institute for Computer Science and Control, Hungary,  ({\tt  email: Y.L.C.Broens@tue.nl}). }
}

\begin{document}

\maketitle
\thispagestyle{empty}
\pagestyle{empty}

\begin{abstract}
The ever increasing need for performance results in increasingly rigorous demands on throughput and positioning accuracy of high-precision motion systems, which  often suffer from position dependent effects that originate from relative actuation and sensing of the moving-body. Due to the highly stiff mechanical design, such systems are typically controlled using rigid body control design approaches. Nonetheless, the presence of position dependent flexible dynamics severely limits attainable position tracking performance.
This paper presents two extensions of the conventional rigid body control framework towards active control of   position dependent flexible dynamics.
Additionally, a novel control design approach is presented, which allows for shaping of the full closed-loop system by means of structured $H_\infty$ co-design. The effectiveness of the approach is validated through simulation using a high-fidelity model of a state-of-the-art moving-magnet planar actuator.
 
\end{abstract}

\section{Introduction}
\label{Section:Introduction}
Increasing production demands in the industry result in a growing need for high-precision and high-throughput mechatronic systems, which are expected to provide complex functionalities while being constantly pressured on their market price. 
Enhanced throughput is obtained by means of lightweight mechanical designs, see \cite{verscheure2006vibration}, which allow for ultra-high accelerations of the moving-body. Nonetheless, the lightweight designs introduce low-frequent resonance dynamics which limit the attainable position tracking accuracy of the mover, see \cite{VibrationModes}. Generally, motion control design of high-precision motion systems is further complicated by position dependent effects, which originate from relative sensing and/or actuation of the moving-body, see \cite{Butler}. To simplify the control design procedure, rigid body coordinate frame transformations are applied to relate the actuation and/or measurement frame to the point of control on the moving body \cite{Murray-book}. However, application of rigid body control design strategies come at the price of introducing position dependent flexible dynamics, see \cite{Steinbuch2013}, forcing the corresponding LTI controller to handle the position dependent flexible modes in terms of robustness at the cost of closed-loop performance.

In recent years, several studies have been conducted which focus on extending the conventional rigid body control framework towards active control of flexible dynamics by means of introducing an output-based modal observer, which allows for reconstruction of flexible modes, see \cite{a059727d0f484157afa82739a695e87e,VOORHOEVE2016642,MaartenSteinbuch202121006010,BAGORDO20116061}. Moreover, the estimated modal signals can be further utilized for active control of flexible dynamics by introducing an additional feedback control loop. Nonetheless, the control design approach presented in \cite{a059727d0f484157afa82739a695e87e} relies on the separations principle for LTI systems, which is conservative in case of position dependent systems.
 
 In order to reduce conservatism, a novel control design approach is presented which allows for simultaneous shaping of the full closed-loop system by means of the structured $H_\infty$ framework, see \cite{apkarian2017h}, which can be trivially extended towards the \emph{linear-parameter-varying} (LPV) framework in the future. Additionally, this paper presents a novel error-based modal observer approach, which reduces the computational complexity of the corresponding control algorithm compared to the conventional output-based modal observer approach.

The main contributions of this paper are:
\begin{itemize}
\item[(C1)] The development of a novel structured $H_\infty$ control co-design approach for a output-based modal observer extension of the well-understood rigid body control design framework, which allows for simultaneous shaping of the full closed-loop system by means of structured $H_\infty$ control synthesis, thus providing local stability and performance guarantees for mechatronic systems that exhibit position dependent effects.
  \item[(C2)] The development of a novel error-based modal observer extension of the rigid body control design structure. Additionally a control co-design approach is developed which allows for synthesis of the full closed-loop system using a single framework.
\end{itemize}

This paper is organized as follows. First, the problem formulation is presented in Section \ref{Section_Problem_formulation}. Next, Section \ref{Section_Outputobserverapproach} presents a structured $H_\infty$ control co-design approach for an output-based modal observer extension of the conventional rigid body framework. Section \ref{Section_Errorobserverapproach}
presents a structured $H_\infty$ control co-design approach for an error-based modal observer extension of the conventional rigid body framework. Section \ref{Section_SimulationStudy} provides a simulation study of the presented approaches on a high-fidelity model of a state-of-the-art moving-magnet planar actuator. Finally, Section \ref{Section_Conclusions} presents the conclusions on the proposed control design approach.

\section{Problem formulation}
\label{Section_Problem_formulation}
\subsection{Background}

To allow for highly accurate positioning of motion systems, sensors and actuators are often physically decoupled from the moving-body, thereby preventing environmental disturbances from affecting positioning accuracy of the mover. Nonetheless, relative actuation and sensing of the moving-body introduce (nonlinear) position dependent effects, which complicate the control design procedure. To efficiently capture these position dependent effects, such systems are often converted to LPV form, see \cite{5714737}. The equations of motion of a multiple-input-multiple-output (MIMO)  mechatronic system in LPV form corresponds to:
\begin{equation}
    M\Ddot{q}(t) + D\dot{q}(t) + Kq(t) = \Phi_\mathrm{a}(p(t)) u(t),
    \label{Section2_Equation_1}
\end{equation}

\noindent where $M$,$D$ and $K$ are the real symmetric mass, damping an stiffness matrices with dimension $n_q \times n_q$ and $\Phi_\mathrm{a}(p(t)) \in \mathbb{R}^{n_q \times n_u}$ maps the input forces $u(t)$ to the appropriate masses based on the scheduling vector $p:\mathbb{R}\rightarrow \mathbb{P}\subseteq\mathbb{R}^{n_p}$ in which the position dependency is embedded. In case that the scheduling vector is constant, implying $p(t) = \tt p \in \mathbb{P}$ for all $t \in \mathbb{R}$, (\ref{Section2_Equation_1}) becomes an LTI system, which is often referred to as  \emph{local dynamics} of a particular LPV system.

In industry, mechatronic systems are typically further transformed to modal form to independently control the mechanical degrees of freedom (DoF), see \cite{gawronski2004dynamics}. The system (\ref{Section2_Equation_1}) is represented in modal form by performing a state transformation $q(t) = \tilde{V}\eta(t)$ using the mass normalized eigenvector matrix $\tilde{V}=M^{-\frac{1}{2}}V$,  which is obtained from the characteristic dynamical equation $KV = MV\Lambda$. The LPV state-space representation of the modal dynamics corresponds to:
\begin{equation}
 G= \left[\begin{array}{cc|c}
      0&I&0\\
      -\Omega^2 & -2Z\Omega & \tilde{V}^\top \Phi_\mathrm{a}(p(t)) \\
      \hline 
     \Phi_\mathrm{s}(p(t))\tilde{V} &0 &0
 \end{array} \right] ,
    \label{Section2_Equation_2}
\end{equation}

\noindent where $Z \in \mathbb{R}^{n_q \times n_q}$ is a diagonal matrix containing the modal damping parameters, $\Omega \in \mathbb{R}^{n_q \times n_q}$  denotes a diagonal matrix containing the eigenfrequencies and $\Phi_\mathrm{s}(p(t))\in \mathbb{R}^{n_y \times n_q}$ maps the position vector $q(t)$ to the output based on the scheduling vector $p(t)$. Furthermore, application of a similarity transformation $T$ to (\ref{Section2_Equation_2}) results in a grouping of the states per mode, where $T$ is denoted by:
\begin{equation}
    T = \left[\begin{array}{cc}
         I_{n_q\times n_q}\otimes \begin{bmatrix}
         1 & 0
         \end{bmatrix}^\top  & I_{n_q \times n_q}\otimes \begin{bmatrix}
         0 & 1
         \end{bmatrix}^\top 
    \end{array} \right],
    \label{Section2_Equation_3}
\end{equation}

\noindent and $\otimes$ corresponds to the \emph{Kronecker-product}. The resulting partitioned state-space representation is given by:
\begin{equation}
    G = \left[\begin{array}{c:c|c}
         A_{\mathrm{RB}} &0 & B_{\mathrm{RB}}(p(t)) \\
         \hdashline
         0 & A_{\mathrm{FM}} & B_{\mathrm{FM}}(p(t)) \\
         \hline 
         C_{\mathrm{RB}}(p(t)) & C_{\mathrm{FM}}(p(t)) & 0
    \end{array} \right],
    \label{Section2_Equation_4}
\end{equation}

\noindent where $\left(\cdot \right)_{\mathrm{RB}}$ are the system matrices that correspond to the rigid body modes and $\left(\cdot \right)_{\mathrm{FM}}$ are the system matrices that coincide with the flexible modes. To allow for loop-shaping based control design techniques, see \cite{skogestad2007multivariable,broens2022lpv,Oomen}, MIMO systems are often decoupled using rigid body (RB) decoupling strategies, see \cite{190Steinbuch}. Rigid body decoupling of the plant is achieved through the decoupling matrices $T_\mathrm{y}$ and $T_\mathrm{u}$, which are typically constructed using:
\begin{equation}
\begin{split}
    T_\mathrm{u} &= \left[\left(I_{n_{\mathrm{RB}} \times n_{\mathrm{RB}}} \otimes \begin{bmatrix}0 & 1 \end{bmatrix} \right)B_{\mathrm{RB}}(p(t))\right]^\dagger \\
    T_\mathrm{y} &= \left[C_{\mathrm{RB}}(p(t))\left(I_{n_{\mathrm{RB}} \times n_{\mathrm{RB}}} \otimes \begin{bmatrix}1 & 0 \end{bmatrix} \right)^\top \right]^\dagger
    \end{split},
    \label{Section2_Equation_6}
\end{equation}

\noindent where $n_{\mathrm{RB}}$ corresponds to the number of rigid body modes of the system. Moreover, the rigid body decoupled system is given by $\tilde{G} = T_\mathrm{y} G T_\mathrm{u}$. Note that introduction of the decoupling matrices $T_{\mathrm{u}}$ and $T_{\mathrm{y}}$ results in elimination of the position dependency in the rigid body dynamics, thus allowing for SISO control design strategies for these mechanical DoFs. Nonetheless, the resonance modes are still coupled in a position dependent manner, therefore introducing limitations regarding achievable position tracking performance. For further improvement of position tracking accuracy of high-precision motion systems, this problem must be addressed.

Our idea is to propose a special (modal) observer that can be co-designed with a modal controller to regulate the remnant effects of the coupled position dependent flexible dynamics to allow for increased closed-loop performance of high-precision motion systems.

\label{Subsection_Background}
\subsection{Problem statement}
\label{Subsection_problem_statement}

The problem that is being addressed in this paper is to extend the industrially well used modal decoupling based $H_\infty$ design to allow for active handling of resonance modes. 
The objective of this paper is to design a structured controller $\mathcal{K}$, such that the following requirements are satisfied.
\begin{itemize}
  \item[(R1)] The closed-loop system is locally stabilized by $\mathcal{K}$ for all $\tt p \in \mathbb{P}$.
  \item[(R2)] The control co-design approach is systematic and  based on optimal gain-based control, such that local closed-loop stability and performance guarantees are obtained for position dependent MIMO systems.
   \item[(R3)] The control design approach is able to impose structure on $\mathcal{K}$, such that physical interpretation of the control algorithm is preserved.
\end{itemize}

\begin{figure}[b]
    \vspace{-15mm}
    \centering
    \includegraphics[trim={0cm 0.5cm 0cm -1cm},width=\linewidth]{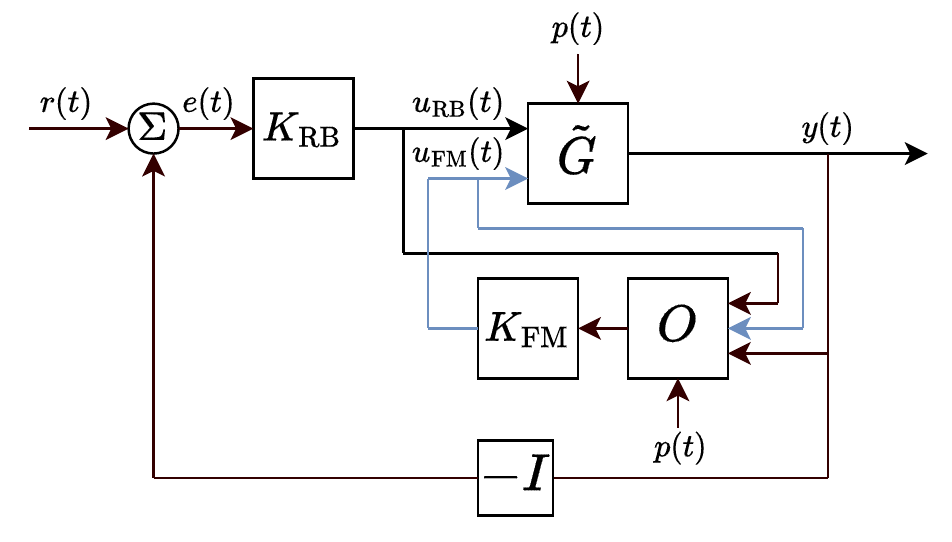}
    \caption{Proposed controller architecture for active control of flexible dynamics using an output-based modal observer.}
    \label{fig:figure1}
\end{figure}


\section{Output-based modal observer approach}
\label{Section_Outputobserverapproach}

This Section presents a novel structured $H_\infty$ control co-design approach for an output-based modal observer extension of the rigid body control framework, which is depicted in Figure \ref{fig:figure1}. To allow for active control of resonance modes, the modal input decoupling is extended towards the flexible modes of interest. Moreover, by considering extended actuator decoupling, the modal decoupled system $\tilde{G}$ is of the form:

\vspace*{-4mm}
\begin{small}
\begin{equation}
    \tilde{G} = T_\mathrm{y} G \left[\left(I_{(n_{\mathrm{RB}}+n_{\mathrm{flex}}) \times (n_{\mathrm{RB}}+n_{\mathrm{flex}})} \otimes \begin{bmatrix}0 & 1 \end{bmatrix} \right)\begin{bmatrix}B_{\mathrm{RB}}(p(t)) \\ 
    B_{\mathrm{FM}}^{n_{\mathrm{flex}}}(p(t))
    \end{bmatrix}\right]^\dagger ,
    \label{modalinputdecoupling}
\end{equation}
\end{small} 

\vspace*{-4mm}
\noindent where $n_{\mathrm{flex}}$ is  the number of flexible modes that are decoupled through the actuation. Extension of the actuator decoupling towards flexible modes introduces several advantageous properties. First, decoupled flexible modes are no longer excited by the feedforward signal due to the orthogonality of excitation directions. Secondly, decoupled flexible modes are no longer visible in the rigid body control loop, thus allowing for increased rigid body feedback control bandwidth. Nonetheless, for fully actuated systems, the extended actuator decoupling introduces channel coupling between the rigid body control loop and the flexible mode control loop. Additionally, decoupled flexible modes still limit attainable position tracking performance as they can be excited by external disturbance sources. Therefore, the rigid body control framework is extended with an output-based modal observer observer $O$, which allows for active control of resonance modes using a flexible mode controller $K_{\mathrm{FM}}$. To account for the introduced channel coupling of fully-actuated systems, a band-pass filter is added to $K_{\mathrm{FM}}$, thereby preventing interaction between the flexible mode control loop and the rigid body control loop.

In order to construct the output-based modal observer $O$, modal truncation is applied to system $\tilde{G}$, which is partitioned as:

\vspace*{-6mm}
\begin{small}
\begin{equation}
    \tilde{G} = \left[\begin{array}{cc:c|c}
         A_{\mathrm{RB}} &0 &0& \tilde{B}_{\mathrm{RB}} \\
         
         0 & A_{\mathrm{FM}}^r &0& \tilde{B}_{\mathrm{FM}}^r \\
         \hdashline 
         0&0&A_{\mathrm{FM}}^d& \tilde{B}_{\mathrm{FM}}^d(p(t))
         \\
         \hline 
         \tilde{C}_{\mathrm{RB}} & \tilde{C}_{\mathrm{FM}}^r(p(t)) &\tilde{C}_{\mathrm{FM}}^d(p(t)) &0
    \end{array} \right],
    \label{PartitionedPlantMOdel}
\end{equation}
\end{small}

\noindent where $\left(\cdot \right)_{\mathrm{FM}}^r$ are the flexible modes that are to be preserved with respect to the output-based modal observer and $\left(\cdot \right)_{\mathrm{FM}}^d$ are the system matrices of the modes that are discarded. Note that due to the extended modal input decoupling (\ref{modalinputdecoupling}), position dependency is removed from $\tilde{B}_{\mathrm{RB}}$, $\tilde{B}_{\mathrm{FM}}^r$ and $\tilde{C}_{\mathrm{RB}}$. The resulting truncated model for observer design, $\hat{G}$, is given by:

\vspace*{-1mm}
\begin{small}
\begin{equation}
    \hat{G} = \left[\begin{array}{cc|c}
         A_{\mathrm{RB}}&0& \tilde{B}_{\mathrm{RB}} \\
         0 & A_{\mathrm{FM}}^r & \tilde{B}_{\mathrm{FM}}^r \\ 
         \hline 
        \tilde{C}_{\mathrm{RB}}  & \tilde{C}_{\mathrm{FM}}^r(p(t)) &-\tilde{C}_{\mathrm{FM}}^d(p(t)){A_{\mathrm{FM}}^d}^{-1}\tilde{B}_{\mathrm{FM}}^d(p(t)) 
    \end{array} \right]
    \label{eq:modaltruncatedmodel}
\end{equation}
\end{small}

\noindent 
where the feed-through matrix corresponds to the compliance correction. Using (\ref{eq:modaltruncatedmodel}), the augmented output-based modal observer $O$ is constructed as:
\begin{equation}
    O:= \left[\begin{array}{c|cc}
         A_o-LC_o(p(t)) & B_o -LD_o(p(t))&L\\
         \hline 
         \Psi & 0
    \end{array}\right],
\end{equation}

\noindent where $ A_o$, $B_o$, $C_o(p(t))$, $D_o(p(t))$ are the state-space matrices of (\ref{eq:modaltruncatedmodel}) and $L$ corresponds to the \emph{Luenberger observer gain}. $\Psi$ is a selection matrix that maps the reconstructed modal velocities to $K_{\mathrm{FM}}$, which is of form:
\begin{equation}
    K_{\mathrm{FM}} = \text{diag}\left(\xi_i \cdot \left(\frac{\frac{\omega_i}{Q}s}{s^2+\frac{\omega_i}{Q}s+\omega_i^2}\right)\right), \quad i \in \left[1 \quad n_{\mathrm{flex}} \right],
    \label{Kfmcontroller}
\end{equation}

\noindent where $\omega_i$ corresponds to the eigenfrequency of the $i$-th controlled flexible mode, $s$ is the complex frequency, $Q$ is a tuning parameter that narrows or broadens the filter band and $\xi_i$ corresponds to a proportional gain.

\begin{figure}[b]
\vspace*{-.4cm}
    \centering
    \includegraphics[trim={0.8cm 0.5cm 0.9cm 0.2cm},width=\linewidth]{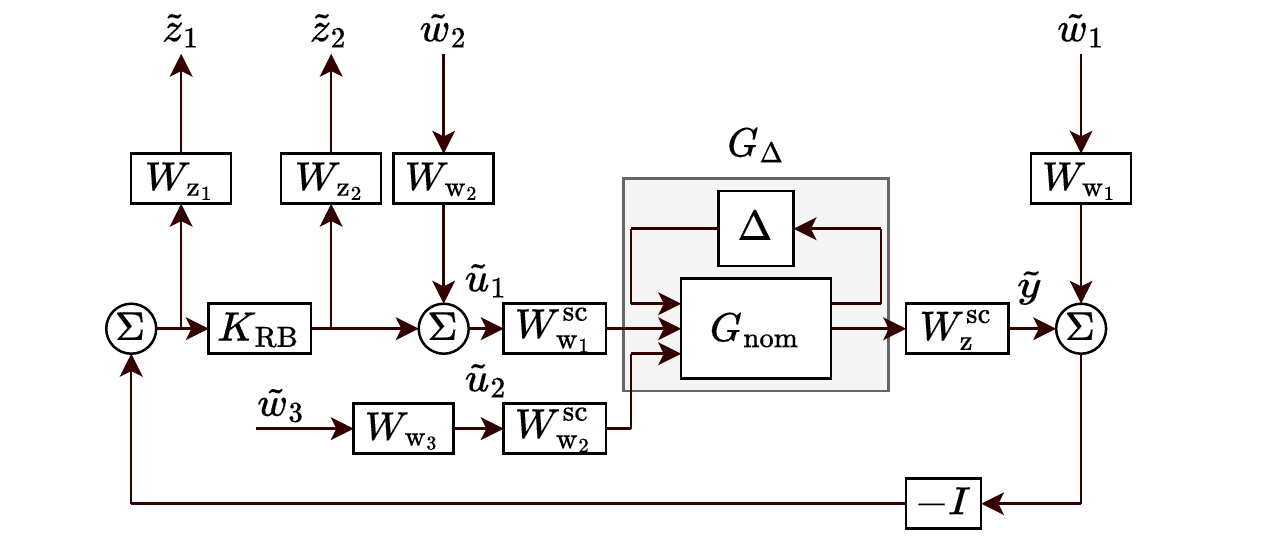}
    \caption{Control interconnection for the structured $H_\infty$ co-design of an output-based modal observer extension of the rigid body control framework.}
    \label{fig:my_label}
\end{figure}

Based on the discussion so far, there is one question yet to be answered. Namely, how to shape the \emph{Luenberger observer gain} $L$, the static state feedback gains $\xi_i$ and the rigid body feedback controller $K_{\mathrm{RB}}$ in an optimal gain-based manner. First, it is observed that the plant is subject to position dependent flexible modes, which can be efficiently modeled as dynamic uncertainties. Moreover, augmentation of the modal observer $O$, the flexible mode controller $K_{\mathrm{FM}}$ and the modal decoupled plant $\tilde{G}$ into an equivalent plant $G_\Delta$, see Figure \ref{fig:my_label}, allows for employment of a novel 6-block control problem (extension of 4-block control problem, see \cite{van2002multivariable}), where $G_{\mathrm{nom}}$ corresponds to the nominal plant and $\left | \left|\Delta \right| \right|_\infty <1$. To allow for appropriate shaping filter design, the plant channels are normalized using the scaling matrices $W_{\mathrm{ z}}^{\mathrm{ sc}}$, $W_{\mathrm{ w_1}}^{\mathrm{ sc}}$ and $W_{\mathrm{ w_2}}^{\mathrm{ sc}}$. Consequently, the scaled uncertain plant $\tilde{G}_\Delta$ is given by:
\vspace*{-1mm}
\begin{equation}
    \tilde{G}_\Delta = W_{\mathrm{ z}}^{\mathrm{sc}} G_\Delta \left[\begin{array}{cc}
         W_{\mathrm{w_1}}^{\mathrm{sc}}& 0 \\
         0& W_{\mathrm{ w_2}}^{\mathrm{sc}}
    \end{array}  \right],
\end{equation}

\noindent where the output scaling filter $W_{\mathrm{z}}^{\mathrm{sc}}$ is chosen as the reciprocals of the expected position tracking error and $W_{\mathrm{w_1}}^{\mathrm{sc}}$ is chosen such that the diagonal elements of $G_\Delta$ have 0 $\mathrm{dB}$ crossings at desired rigid body target bandwidths $f_{\mathrm{bw}}^i$, where $i\in \begin{bmatrix}1& n_{\mathrm{RB}} \end{bmatrix}$. Moreover, the input scaling filter $W_{\mathrm{w_1}}^{\mathrm{sc}}$ is chosen as:
\begin{equation}
    W_{\mathrm{w_1}}^{\mathrm{sc}} = \text{diag}\left(\text{abs}(G_{\mathrm{nom}}(j(2\pi f_{\mathrm{bw}}^i))) \right)^{-1} {W_{\mathrm{z}}^{\mathrm{sc}}}^{-1}
\end{equation}

\noindent while the other scaling matrix is set to $W_{\mathrm{w_2}}^{\mathrm{sc}} = I$. Furthermore, the closed-loop transfer between the generalized inputs and the generalized outputs, see Figure \ref{fig:my_label}, is given by:
\vspace*{-1mm}
\begin{equation}
    \left[\begin{array}{c}
         \tilde{z}_1 \\ \tilde{z}_2 
    \end{array} \right] = 
    -M
    \left[\begin{array}{c}
         \tilde{w}_1 \\ \tilde{w}_2 \\ \tilde{w}_3
    \end{array} \right],
\end{equation}

\vspace*{-1mm}
\noindent where $M$ corresponds to:
\begin{equation}
\resizebox{.88\hsize}{!}{$
    \left[\begin{array}{cc:c}
      W_{\mathrm{z_1}}S W_{\mathrm{w_1}}   &  W_{\mathrm{z_1}}S\tilde{G}_{\Delta_{\tilde{y},\tilde{u}_1}} W_{\mathrm{w_2}} & W_{\mathrm{z_1}}S\tilde{G}_{\Delta_{\tilde{y},\tilde{u}_2}} W_{\mathrm{w_3}} \\
       W_{\mathrm{z_2}}K_{\mathrm{RB}}S W_{\mathrm{w_1}} & W_{\mathrm{z_2}}K_{\mathrm{RB}}S\tilde{G}_{\Delta_{\tilde{y},\tilde{u}_1}} W_{\mathrm{w_2}} & W_{\mathrm{z_2}}K_{\mathrm{RB}}S\tilde{G}_{\Delta_{\tilde{y},\tilde{u}_2}} W_{\mathrm{w_3}}
    \end{array} \right]$},
    \label{Extended4Block}
\end{equation}

\noindent and $S = [I+\tilde{G}_{\Delta_{\tilde{y},\tilde{u}_1}}K_{\mathrm{RB}}]^{-1}$. Therefore, the controller synthesis objective is given by:
\begin{equation}
  \min_{\mathcal{K}}  \left| \left|M \right| \right|_{\infty}, \quad  \text{where }\mathcal{K} = \text{diag}\left(K_{\mathrm{RB}},L,K_{\mathrm{FM}}\right).
  \label{optimizationobjective}
\end{equation}

From a design perspective, it is observed from (\ref{Extended4Block}) that the proposed control design approach allows for simultaneous shaping of the full rigid body control loop and the flexible mode control loop. In order co-design the controller in a desirable manner, it is important to encode desirable controller properties through the shaping filters $W_{\mathrm{z_1}}, W_{\mathrm{z_2}}, W_{\mathrm{w_1}}$, $W_{\mathrm{w_2}}$ and $W_{\mathrm{w_3}}$, such as integral action, roll-off action, appropriate rigid body feedback control bandwidth and active damping of resonance modes.
\subsection{Integral action}

From the optimization objective denoted by (\ref{optimizationobjective}), it is observed that integral action is enforced on $K_{\mathrm{RB}}$ by imposing a +20 $\frac{\mathrm{dB}}{\mathrm{dec}}$ slope on the normalized process sensitivity for low-frequencies. This is achieved through shaping filter $W_{\mathrm{z_1}}$, which is of form:
\begin{equation}
    W_{\mathrm{z_1}} = \text{diag}\left(K_s \frac{s+2\pi f_{I_i}}{s} \right), \quad i \in \begin{bmatrix}
    1&n_{\mathrm{RB}}
    \end{bmatrix},
    \label{shapingwz1}
\end{equation}

\noindent where integral action is enforced on $K_{\mathrm{RB}}$ up  and until the cut-off frequency $ f_{I_i}\approx \frac{1}{4}f_{\mathrm{ bw}}^i$ in order to suppress low-frequent disturbances. Typically, $K_s$ is chosen to be 0.5 in to place an upper-bound of 6 $\mathrm{dB}$ on the sensitivity and the normalized process sensitivity. As the sensitivity is already shaped in a desirable manner, $W_{\mathrm{w_1}}$ can be chosen to be $I$. 
\subsection{Roll-off action}

To suppress high-frequent disturbances, roll-off action is imposed on $K_{\mathrm{RB}}$ by enforcing a -20 $\frac{\mathrm{dB}}{\mathrm{dec}}$ slope on the complementary sensitivity and the normalized control sensitivity for high frequencies. This is realized by shaping $W_{\mathrm{z_2}}$ as:
\begin{equation}
    W_{\mathrm{z_2}} = \text{diag}\left(K_r \frac{s+2\pi f_{r_i}}{\frac{1}{\alpha}s+2\pi f_{r_i}} \right),
    \quad i \in \begin{bmatrix}
    1&n_{\mathrm{RB}}
    \end{bmatrix},
    \label{dezefunctie}
\end{equation}

\noindent where $K_r$ is set to 0.5 in order to place an upper-bound of 6 $\mathrm{dB}$ on the normalized control sensitivity and the complementary sensitivity. Additionally, the cut-off frequency of the roll-off action is typically taken as $f_{r_i} \approx 4 f_{\mathrm{ bw}}^i$ with $\alpha = 20$ to ensure that the shaping filter proper and to guarantee satisfactory rigid body feedback control bandwidth. Additionally, shaping filter $W_{\mathrm{ w_2}}$ is used to impose additional roll-off action or notch action on $K_{\mathrm{RB}}$ if desired. Otherwise $W_{\mathrm{w_2}}=I$. 

\subsection{Active damping of resonance modes}

For co-design of the flexible mode controller, shaping filter $W_{\mathrm{ w_3}}$ is designed to inflict damping on the flexible mode control loop to synthesize the proportional gains $\xi_i$ and the observer gain $L$. This is realized by choosing $W_{\mathrm{w_3}}$ to be of form:

\begin{small}
\begin{equation}
    W_{\mathrm{w_3}} = \mathrm{diag}\left(\epsilon_i \cdot  \frac{\frac{1}{(2\pi f_1^i)^2} s^2+\frac{2\beta_1^i}{2\pi f_1^i}s+1}{\frac{1}{(2\pi f_2^i)^2} s^2+\frac{2\beta_2^i}{2\pi f_2^i}s+1}\right), \quad i \in \begin{bmatrix}1&n_{\mathrm{flex}} \end{bmatrix},
    \label{notchfilter}
\end{equation}
\end{small}

\noindent where $f_1$ = $f_2$ is the eigenfrequency of the $i$-th flexible mode, $\beta_1^i > \beta_2^i$ and $\epsilon_i$ is a tuning parameter which allows to shape the \emph{flexible mode process sensitivity}, see (\ref{Extended4Block}), such that desirable damping is achieved, where $L$ is initialized in the structured $H_\infty$ optimization with the solution of the algebraic Riccati equation.

In total, a robust control design approach is presented for co-design of the full closed-loop system of an output-based modal observer extension of the rigid body control loop. Note that the presented control approach can be extended towards LPV synthesis to achieve increased closed-loop performance.

\section{Error-based modal observer approach}
\label{Section_Errorobserverapproach}
\begin{figure}[h]
    \centering
    \includegraphics[trim={0cm 0.5cm 0cm 0.4cm},width=\linewidth]{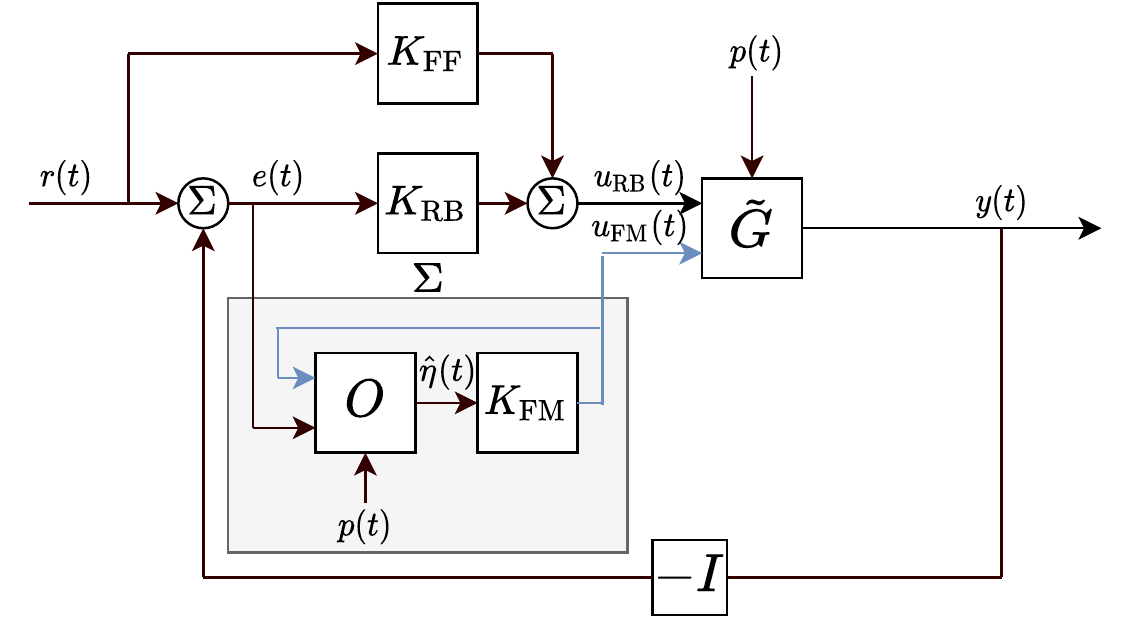}
    \caption{Proposed controller architecture for active control of flexible dynamics using an error-based modal observer.}
    \label{fig:errorobserver}
\end{figure}

This Section presents a novel structured $H_\infty$ control co-design approach for an error-based modal observer extension of the rigid body control framework, which is depicted in Figure \ref{fig:errorobserver}. 
Note that under the assumption that the rigid body feedforward perfectly cancels out the rigid body modes from the error signal, implying $e(t) \approx -y_{\mathrm{flex}}(t)$, the observer dynamics can be solely constructed from flexible modes, thus reducing the state order of the modal observer. Additionally, due to the modal input decoupling, the rigid body control forces do not affect the flexible modes, therefore reducing the input mapping of the corresponding error-based observer.


In order to construct the model for observer design, consider the partitioned plant model expressed by equation (\ref{PartitionedPlantMOdel}), where we only consider preservation of flexible modes.
Note that since we are observing the error $e(t)$ based on the assumption that the rigid body feedforward perfectly cancels out the rigid body modes, the output equation is adjusted accordingly, i.e. $e(t) \approx -y_{\mathrm{flex}}(t)$. Moreover, the error-based modal observer $O$ is given by: 
\begin{equation}
    O := \left[\begin{array}{c|cc}
         A_{\mathrm{FM}}^r + L \tilde{C}_{\mathrm{FM}}^r (p(t))& \tilde{B}_{\mathrm{FM}}^r + L D_o(p(t)) &L\\ \hline \Psi &0 &0
    \end{array} \right] ,
    \label{Modalerrorobserver}
\end{equation}


To construct a structured $H_\infty$ control co-design approach, subsystem $\Sigma$ is introduced, see Figure \ref{fig:errorobserver}, where $\Sigma$ corresponds to:
\vspace*{-3mm}
\begin{equation}
    \Sigma = \left[I-K_{\mathrm{FM}}O_{\hat{\eta},u_{\mathrm{FM}}} \right]^{-1}K_{\mathrm{FM}}O_{\hat{\eta},e},
\end{equation}

\vspace*{-1mm}
\noindent where $O_{\hat{\eta},u_{\mathrm{FM}}}$ denotes the observer mapping from $u_{\mathrm{FM}}(t)$ to $\hat{\eta}(t)$ and $O_{\hat{\eta},e}$ is the mapping from $e(t)$ to $\hat{\eta}(t)$.

\begin{figure}[h]
    \centering
    \includegraphics[trim={0.8cm 0.5cm 0.25cm 0cm},height = 3.5cm,width=.8\linewidth]{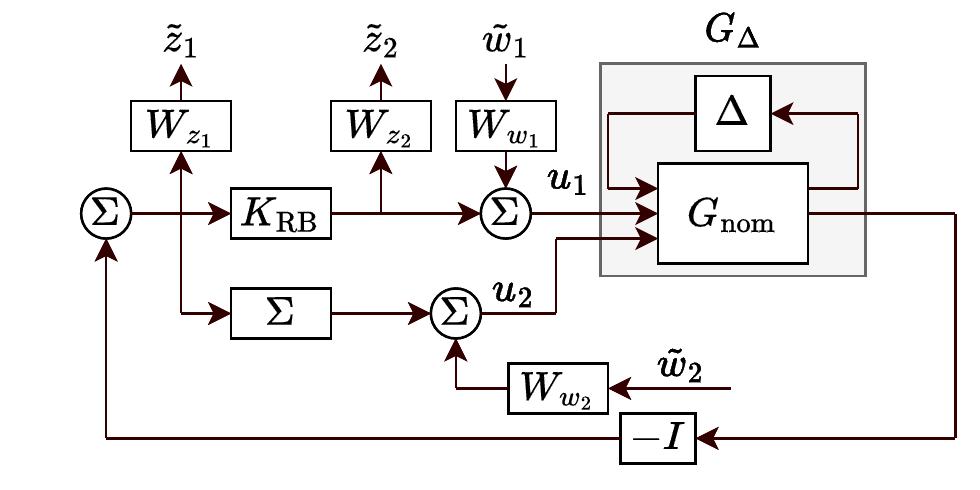}
    \caption{Control interconnection for the structured $H_\infty$ co-design of an error-based modal observer extension of the rigid body control framework.}
    \label{fig:2blocksynthesis}
\end{figure}

For controller synthesis, the rigid body feedforward is eliminated from the control design by reformulating the design problem into a disturbance rejection problem, see \cite{skogestad2007multivariable}. Additionally, a secondary generalized input channel $\tilde{w}_2$ is introduced to allow for active shaping of the flexible mode controller $K_{\mathrm{FM}}$ and the observer gain $L$. The corresponding control interconnection is illustrated by Figure \ref{fig:2blocksynthesis}, where the closed-loop transfer between the generalized inputs and the

\newpage 

\noindent 
generalized outputs is given by:

\begin{small}
\begin{equation}
    \left[\begin{array}{c}
         \tilde{z}_1 \\ \tilde{z}_2
    \end{array} \right] = 
   - M \left[\begin{array}{c}
         \tilde{w}_1 \\ \tilde{w}_2
    \end{array} \right],
    \label{2blocksynthesis}
\end{equation}
\end{small}

\noindent where $M$ corresponds to:
\begin{equation}
     \left[\begin{array}{c:c}
    W_{\mathrm{z_1}}\tilde{S} G{_\Delta{_{y,u_1}}} W_{\mathrm{w_1}} 
    &
   W_{\mathrm{z_1}}\tilde{S} G{_\Delta{_{y,u_2}}} W_{\mathrm{w_2}} 
    \\
    W_{\mathrm{z_2}}K_{\mathrm{RB}}\tilde{S} G{_\Delta{_{y,u_1}}} W_{\mathrm{w_1}} &
    W_{\mathrm{z_2}}K_{\mathrm{RB}}\tilde{S} G{_\Delta{_{y,u_2}}} W_{\mathrm{w_2}}
    \end{array} \right]
\end{equation}

\noindent and $\tilde{S} = \left[I+G{_\Delta{_{y,u_1}}}K_{\mathrm{RB}}+G{_\Delta{_{y,u_2}}}\Sigma \right]^{-1}$. Therefore, the synthesis objective is given by:
\begin{small}
\begin{equation}
    \min_{\mathcal{K}} \left| \left|\left[\begin{array}{cc}
    W_{\mathrm{z_1}}\tilde{S} G{_\Delta{_{y,u_1}}} W_{\mathrm{w_1}} 
    &
   W_{\mathrm{z_1}}\tilde{S} G{_\Delta{_{y,u_2}}} W_{\mathrm{w_2}} 
    \\
    W_{\mathrm{z_2}}K_{\mathrm{RB}}\tilde{S} G{_\Delta{_{y,u_1}}} W_{\mathrm{w_1}} &
    W_{\mathrm{z_2}}K_{\mathrm{RB}}\tilde{S} G{_\Delta{_{y,u_2}}} W_{\mathrm{w_2}}
    \end{array} \right] \right| \right|_\infty , 
    \label{Onjective2block}
\end{equation}
\end{small}

\vspace*{-3mm}
\noindent where $\mathcal{K} = \text{diag}\left(K_{\mathrm{RB}},L,K_{\mathrm{FM}} \right)$. Note that the left column allows for the shaping of the rigid body process sensitivity and the complementary sensitivity, while the right column allows for the shaping of the flexible mode control loop.

Similarly to the design approach that was presented in Section \ref{Section_Outputobserverapproach}, desirable controller properties, such as integral action, roll-off action and active damping, are encoded through the appropriate shaping filters.
\subsection{Integral action}
Integral action is imposed by shaping filter $W_{\mathrm{z_1}}$, which is of similar form as (\ref{shapingwz1}). Nonetheless, the proportional gain coefficients $K_s$ are chosen such that the diagonal elements of the process sensitivity shaping filter (e.g. $[W_{\mathrm{z_1}} W_{\mathrm{w_1}}]^{-1}$) are upper-bounded by a desirable magnitude in order to provide sufficient disturbance rejection.
\subsection{Roll-off action}
From (\ref{Onjective2block}) it is observed that $W_{\mathrm{w_1}}$ both shapes the rigid body process sensitivity and the rigid body complementary sensitivity. Due to the desire to encode roll-off action in $K_{\mathrm{RB}}$, $W_{\mathrm{w_1}}$ is shaped as (\ref{dezefunctie}). Next,  $W_{\mathrm{z_2}}=I$ since $W_{\mathrm{w_1}}$ already imposes a desirable structure on the rigid body complementary sensitivity.
\subsection{Active damping of resonance modes}
Shaping of the flexible mode loop is realized through shaping filter $W_{\mathrm{w_2}}$, which is of similar form as (\ref{notchfilter}), where $\epsilon_i$ allows for scaling of the closed loop relations of the flexible mode control loop, such that desirable closed-loop behaviour is obtained.

In total, a novel extension of the rigid body control framework is presented by means of an error-based modal

\noindent 
observer, which reduces the required state order of the flexible mode control loop as no rigid body states are required. Additionally, a structured $H_\infty$ design approach has been presented, which allows for co-design of the rigid body feedback controller $K_{\mathrm{RB}}$, the flexible mode controller $K_{\mathrm{FM}}$ and the observer gain $L$, which is initialized with the solution of the algebraic Riccati equation.

\section{Simulation study}
\label{Section_SimulationStudy}

\begin{figure}[b]
\vspace*{-3mm}
    \centering
    \includegraphics[trim={0cm 0cm 0.5cm 0cm},width=.9\linewidth]{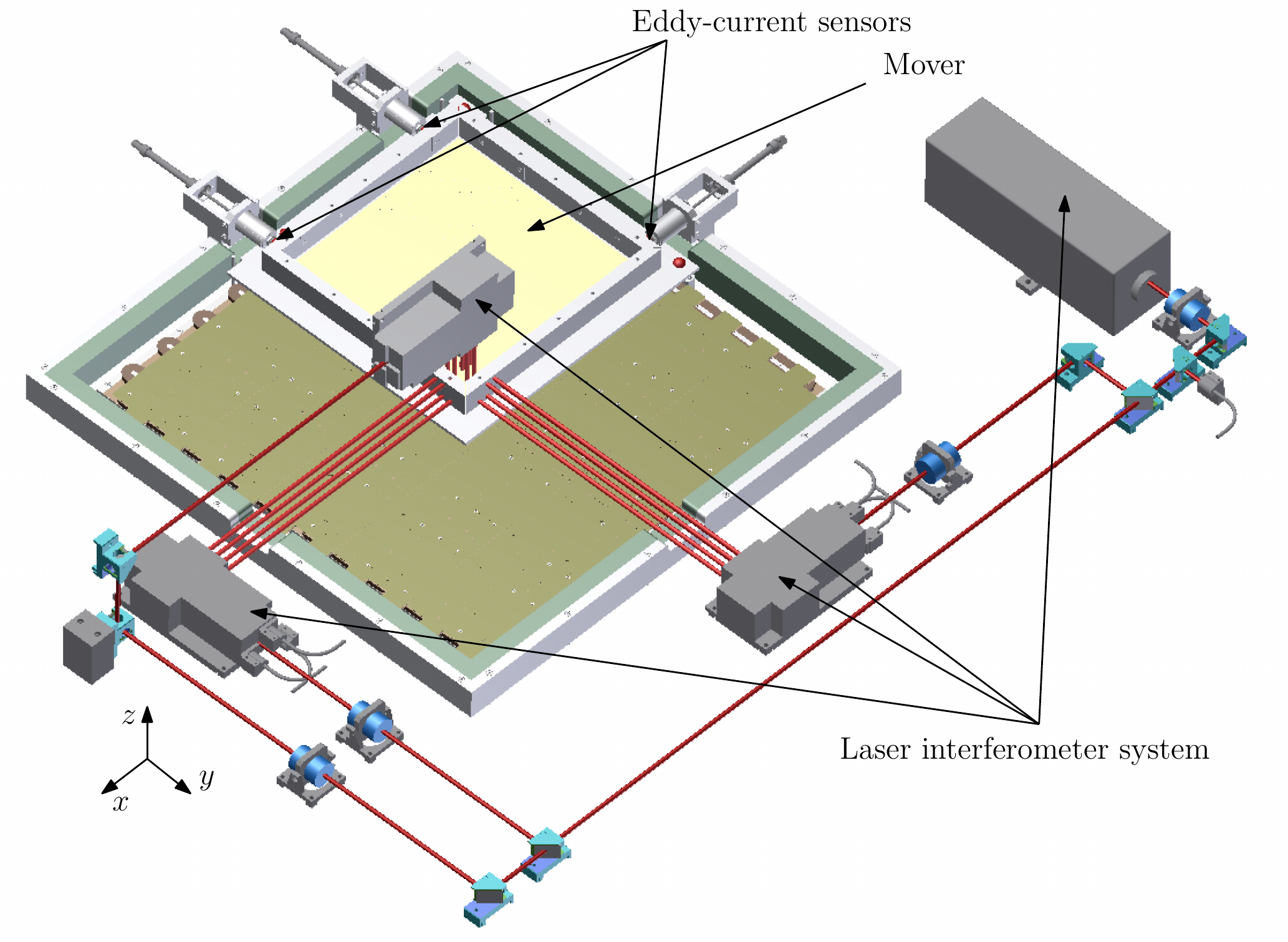}
    \vspace*{-4mm}
    \caption{Schematic representation of a MMPA prototype.}
    \label{fig:NAPAS}
\end{figure}

This section presents a simulation study of the proposed control design approaches using a high-fidelity model of a moving-magnet planar actuator (MMPA) system, which is depicted in Figure \ref{fig:NAPAS}. A MMPA is a high-precision motion system which exhibits position dependent flexible dynamics due to position dependent actuation and sensing of the mover.
For a detailed description of such a system, see \cite{9566789}.

\subsection{Output-based modal observer approach}
\label{subsectionoutputobserver}

\begin{figure}[t]
    \centering
    \includegraphics[trim={4cm .8cm 3.5cm 1cm},clip,width=\linewidth,width=\linewidth]{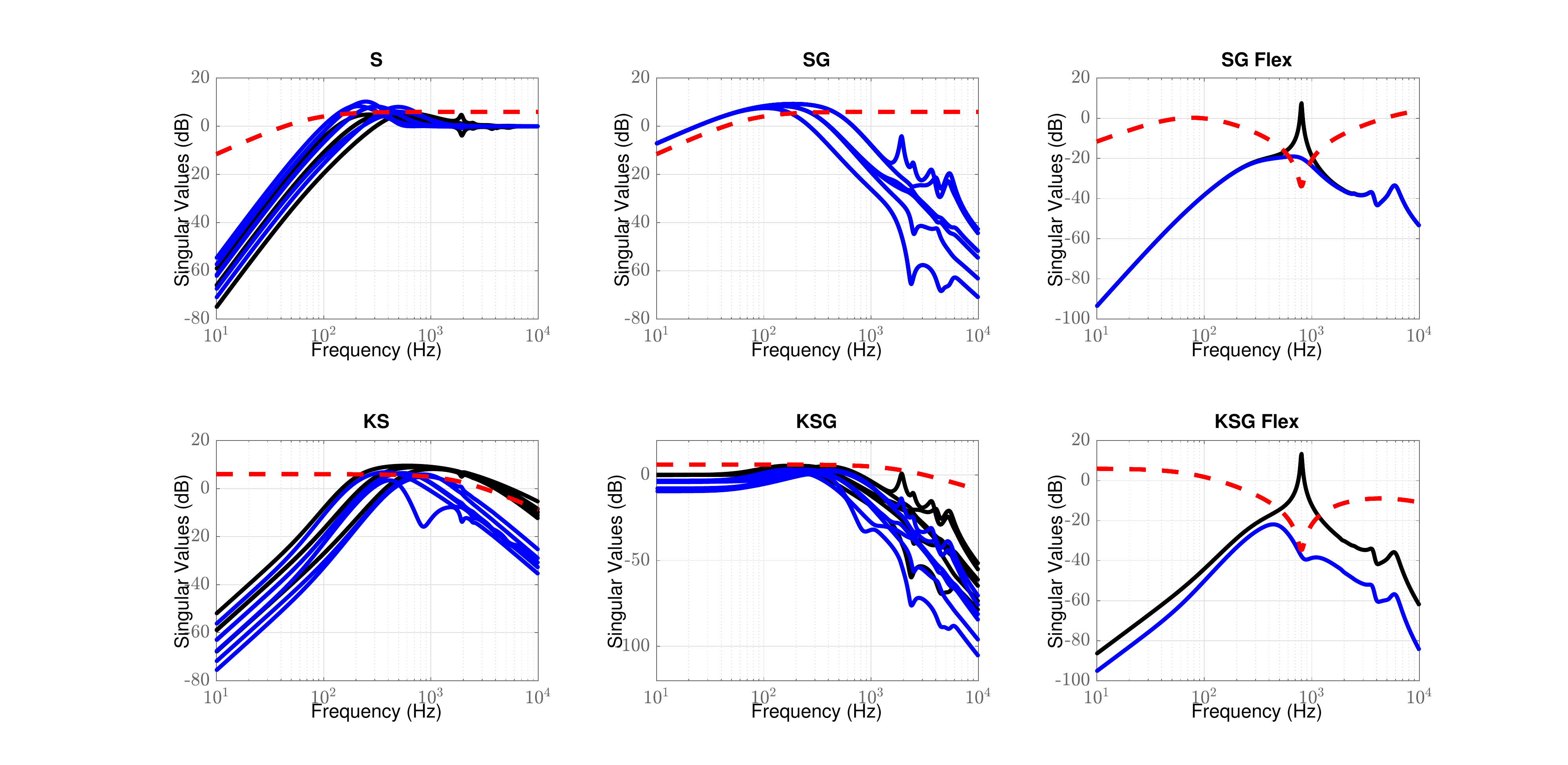}
    \caption{Structured $H_\infty$ synthesis results, with (\Large\textcolor{red}{-}\footnotesize): shaping filters, (\Large\textcolor{black}{-}\footnotesize): conventional rigid body control approach, (\Large\textcolor{blue}{-}\footnotesize): proposed control approach.}
    \label{fig:6blockresult}
    \vspace*{-6mm}
\end{figure}

For the structured $H_\infty$ synthesis of the output-based modal observer extension,  the 6-block synthesis, expressed by (\ref{Extended4Block}), is considered. For the conventional rigid body control approach, 4-block shaping is considered (see \cite{van2002multivariable}), where for comparison of the two control approaches, equivalent shaping filters are used. The synthesis results are illustrated in Figure \ref{fig:6blockresult}, where the dotted red graph corresponds to the shaping filters, the black graph denotes the conventional rigid body control structure (4-block) and the blue graph coincides 

\noindent 
with the scenario for which the output-based modal observer is active (6-block). From Figure \ref{fig:6blockresult}, several observations can be made. First, it is clearly visible from the third column that introduction of the output-based modal observer to the rigid body control structure results in damping of the flexible mode.  Additionally, it is concluded that both control design

\noindent   approaches achieve similar rigid body feedback control bandwidth, which can be explained by the extended modal input decoupling which ensures that decoupled flexible modes are no longer visible in the rigid body feedback loop, thus allowing for increased rigid body feedback control bandwidth. Moreover, the proposed control design approach allows for increased closed-loop performance compared to conventional rigid body control approaches due to the active control of resonance modes.

\subsection{Error-based modal observer approach}
\label{subsectionerrorobserver}

\begin{figure}[b]
    \vspace*{-6mm}
    \centering
    \includegraphics[trim={4cm 1.2cm 3.5cm 1cm},clip,width=\linewidth]{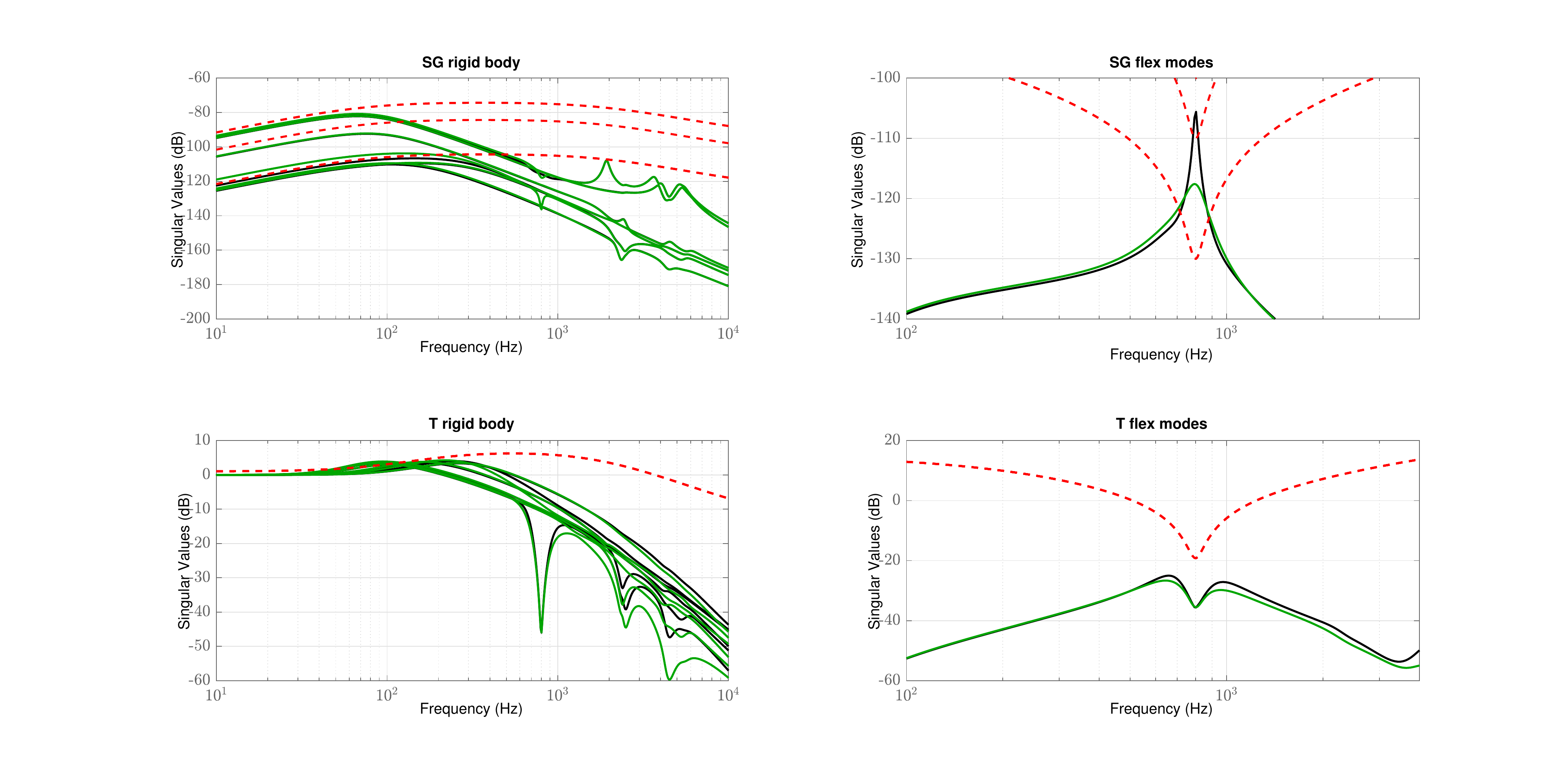}
    \caption{Structured $H_\infty$ synthesis results, with (\Large\textcolor{red}{-}\footnotesize): shaping filters, (\Large\textcolor{black}{-}\footnotesize): conventional rigid body control approach, (\Large\textcolor{green}{-}\footnotesize): proposed control approach.}
    \label{fig:errorobs}
\end{figure}

For the structured $H_\infty$ synthesis of the error-based modal observer, the 4-block synthesis, expressed by (\ref{Onjective2block}), is considered.
Similar to the comparison in Subsection \ref{subsectionoutputobserver}, two scenarios are investigated. Namely, when the error-based modal observer loop is active and when the observer loop is inactive. The synthesis results are illustrated in Figure \ref{fig:errorobs}, where the red dotted graph corresponds to the shaping filters, the black graph  is the scenario for which the modal observer loop is inactive and the green graph corresponds to the scenario for which the error-based modal observer is active. From the Figure, several observations are made. First, it is observed from the control interconnection illustrated by Figure \ref{fig:2blocksynthesis} that the synthesis of the rigid body feedback controller and the flexible mode controller is coupled due to their dependency on the position tracking error signal $e(t)$. Nonetheless, from the synthesis results, illustrated by Figure \ref{fig:errorobs}, it is clearly visible from the right column that active damping is achieved by introducing the error-based modal observer to the rigid body control structure. However, using the error-based modal observer extension of the rigid body feedback control framework, a trade-off is introduced between the performance of the rigid body feedback control loop (e.g. left column of Figure \ref{fig:2blocksynthesis}) and the performance of the suppression of the flexible mode (right column of Figure \ref{fig:2blocksynthesis}) due to the coupling of both control loops. Therefore, achievable closed-loop performance of the error-based modal observer is limited compared to the output-based modal observer structure.



\section{Conclusions}
\label{Section_Conclusions}
This paper presents a novel structured $H_\infty$ control co-design approach for two modal observer based extensions of the conventional rigid body control framework, which allows for co-design of the full closed-loop system by utilizing extended input decoupling. Specifically, a novel error-based modal observer structure is presented, which is attractive from a computational point of view. Nonetheless, the reduction in complexity comes at the cost of closed-loop performance. The proposed approach allows for extension towards the linear-parameter-varying control design framework, such that closed-loop performance of high-precision position dependent motion systems can be pushed beyond the reach of LTI control design strategies.

\addtolength{\textheight}{-12cm}   




\bibliographystyle{ieeetr}       
\bibliography{MyBib}

\end{document}